\documentclass[prd,twocolumn,aps,showpacs,nofootinbib,nobibnotes,superscriptaddress,preprintnumbers]{revtex4}
\usepackage{epsfig}
\usepackage{graphics}
\usepackage{bm}
\usepackage{color}
\usepackage{dcolumn}   
\usepackage{bm}     
\usepackage{bbm}       
\usepackage{amssymb}  
\usepackage{amsmath}
\usepackage{latexsym}
\usepackage{float}
\usepackage{ifthen}
\usepackage{caption,subfig}
\usepackage{enumerate}
\usepackage{url}
\usepackage{caption,subfig}
\usepackage{amsopn}

\usepackage{amsfonts}
\usepackage{multirow}
\usepackage{array}
\usepackage{booktabs}
\usepackage{rotating}

\usepackage{ulem}
\newcommand{\ud}{\mathrm{d}} 

\def\clap#1{\hbox to 0pt{\hss#1\hss}}

\def\({\left(}
\def\){\right)}
\def\[{\left[}
\def\]{\right]}
\def\bea{\begin{eqnarray}}
\def\eea{\end{eqnarray}}
\def\be{\begin{equation}}
\def\ee{\end{equation}}
\def\ba{\begin{eqnarray}}
\def\ea{\end{eqnarray}}
\def\beq{\begin{eqnarray}}
\def\eeq{\end{eqnarray}}
\def\mpl{M_{\rm P}}

\newcommand{\cs}{c_s}

\newcommand{\Id}{\mathbbm 1}

\def\cs{c_{\rm s}}

\def\be{\begin{equation}}
\def\ee{\end{equation}}
\def\ba{\begin{eqnarray}}
\def\ea{\end{eqnarray}}
\def\beq{\begin{eqnarray}}
\def\eeq{\end{eqnarray}}

\def\mpl{M_{\rm P}}

\def\L*{{\cal L}_*}
\def\L{\mathcal{L}}
\def\({\left(}
\def\){\right)}

\def\<{\langle}
\def\>{\rangle}

\def\cs2{c_{s}^{2}}

\def\be{\begin{equation}}
\def\ee{\end{equation}}
\def\ba{\begin{eqnarray}}
\def\ea{\end{eqnarray}}
\def\beq{\begin{eqnarray}}
\def\eeq{\end{eqnarray}}

\def\mpl{M_{\rm P}}

\def\L*{{\cal L}_*}
\def\L{\mathcal{L}}
\def\({\left(}
\def\){\right)}

\def\<{\langle}
\def\>{\rangle}

\begin{document}
\hspace{5.2in} \mbox{NORDITA-2015-38}\\\vspace{1.53cm} 

\title{Dark Matter via Massive (bi-)Gravity}

\date{\today,~ $ $}

\author{Luc Blanchet} \email{blanchet@iap.fr}
\affiliation{$\mathcal{G}\mathbb{R}\varepsilon{\mathbb{C}}\mathcal{O}$
  Institut d'Astrophysique de Paris --- UMR 7095 du CNRS,
  \ Universit\'e Pierre \& Marie Curie, 98\textsuperscript{bis}
  boulevard Arago, 75014 Paris, France}

\author{Lavinia Heisenberg} \email{laviniah@kth.se}
\affiliation{Nordita, KTH Royal Institute of Technology and Stockholm
  University, \\Roslagstullsbacken 23, 10691 Stockholm, Sweden}
\affiliation{Department of Physics \& The Oskar Klein Centre, AlbaNova
  University Centre, 10691 Stockholm, Sweden}

\date{\today}

\begin{abstract}
  In this work we investigate the existence of relativistic models for
  dark matter in the context of bimetric gravity, used here to
  reproduce the modified Newtonian dynamics (MOND) at galactic
  scales. For this purpose we consider two different species of dark
  matter particles that separately couple to the two metrics of
  bigravity. These two sectors are linked together \textit{via} an
  internal $U(1)$ vector field, and some effective composite metric
  built out of the two metrics. Among possible models only certain
  classes of kinetic and interaction terms are allowed without
  invoking ghost degrees of freedom. Along these lines we explore the
  number of allowed kinetic terms in the theory and point out the
  presence of ghosts in a previous model. Finally, we propose a
  promising class of ghost-free candidate theories that could provide
  the MOND phenomenology at galactic scales while reproducing the
  standard cold dark matter (CDM) model at cosmological scales.
\end{abstract}


\maketitle

\section{Introduction}

General Relativity (GR) successfully describes the gravitational
interaction in a wide range of scales and regimes, from the solar
system size to strong fields in binary pulsars and black holes, and
most likely will constitute the correct tool for the future
gravitational wave astronomy~\cite{Will}. Up to now, GR has been able
to prevail against all alternative theories, either
scalar-tensor~\cite{Horndeski:1974wa, Nicolis:2008in,Deffayet:2009wt,
Deffayet:2009mn, deRham:2011by, Heisenberg:2014kea}, 
vector-tensor~\cite{Horndeski:1976gi,EPU10,PhysRevD.80.023004,
BeltranJimenez:2013fca,Jimenez:2013qsa,Heisenberg:2014rta,
Tasinato:2014eka} or tensor-tensor theories, the latter comprising
massive gravity~\cite{deRham10, dRGT10}, bigravity~\cite{Hassan12a,
  Hassan12b} and multigravity~\cite{Hinter12} theories.

In spite of these successes, the extrapolation of GR to a broader
range of scales --- notably, cosmological scales --- faces important
challenges since it relies on the introduction of a dark sector,
composed of dark matter and dark energy. The nature of this dark
sector constitutes one of the most important mystery of contemporary
physics.

The reference model of cosmology today assumes a pure cosmological
constant $\Lambda$ added to the field equations of GR to account for
the dark energy, and a component of non-baryonic dark matter made of
non relativistic particles called cold dark matter (CDM). The best
motivated candidate for the dark matter particle is the
WIMP~\cite{BHS05}. The model $\Lambda$-CDM is very well tested at
cosmological scales by the accelerated expansion of the universe, by
the observed fluctuations of the cosmic microwave background, and by
the distribution of dark matter in large scale structures.

Unfortunately this model does not explain the presence of a tiny
cosmological constant $\Lambda$. In the prevailing view it should be
interpreted as a constant energy density of the vacuum. However, the
unnatural observed value of $\Lambda$ and the instability against
large quantum corrections put in doubt its consistency using standard
quantum field theory techniques.

Another important concern is that the model $\Lambda$-CDM does not
account for many observations of dark matter at the scale of galaxies,
where it faces unexplained tight correlations between dark and
luminous matter in galaxy halos~\cite{SandMcG02,
  FamMcG12}. Primary examples are the baryonic Tully-Fisher
  relation between the asymptotic rotation velocity of spiral galaxies
  and their baryonic mass, and the correlation between the mass
  discrepancy (\textit{i.e.} the presence of dark matter) and the
  acceleration scale involved~\cite{McG00, McG11}. These correlations
  happen to be very well explained by the MOND (MOdified Newtonian
  Dynamics) empirical formula~\cite{Milg1, Milg2, Milg3}. The
agreement between MOND and all observations at galactic scales is
remarkable and calls for an explanation. On the other hand, MOND has
problems explaining the DM distribution at the larger scale of galaxy
clusters~\cite{GD92, PSilk05, Clowe06, Ang08, Ang09}.

Many works have been devoted to promoting the MOND formula into a
decent relativistic theory. Most approaches modify GR with extra
fields without invoking dark matter~\cite{Sand97, Bek04, Sand05,
  ZFS07, Halle08, bimond1, BDgef11, BM11, Arraut2014}. Here we shall
be interested in another approach, based on a form of dark matter
\textit{\`a la} MOND called dipolar dark matter (DDM). This approach
is motivated by the dielectric analogy of MOND~\cite{B07mond}. A first
relativistic model was proposed in~\cite{BL08, BL09} and shown to
reproduce the model $\Lambda$-CDM at cosmological scales. Recently, a
more sophisticated model has been based on a bimetric extension of
GR~\cite{BB14} (see also~\cite{BBwag} for further motivation). In this
model two species of dark matter particles are coupled respectively to
the two metrics, and are linked by an internal vector field generated
by the mass of these particles. The phenomenology of MOND then results
from a mechanism of gravitational polarization.

Bimetric theories have been extensively investigated in the quest of a
consistent massive gravity theory going beyond the linear Fierz-Pauli
theory. The past decade has seen the emergence of a specific
theory~\cite{deRham10, dRGT10} that avoids the appearance of the
Boulware-Deser (BD) ghost~\cite{BoulwareD} to any order in
perturbations. This dRGT theory~\cite{deRham10, dRGT10} has been
extended and reformulated as a bimetric theory with two dynamical
metrics~\cite{Hassan12a, Hassan12b}. The theoretical and cosmological
implications of these theories are extremely rich. Notably,
cosmological solutions of massive gravity theories have drawn much
attention~\cite{deRham:2010tw,PhysRevD.84.124046,PhysRevLett.109.171101}
(see also the references in~\cite{deRham:2014zqa}).

In the present paper we point out that the previous model for DDM in a
bimetric context~\cite{BB14}, despite the important phenomenology it
is able to reproduce, is plagued by ghosts and cannot be considered as
a viable theory. Nevertheless, this phenomenology (especially at
galactic scales, \textit{i.e.} MOND) definitely calls for a more
fundamental theory. We look for a consistent coupling of the dark
matter fields to bigravity, closely following the restrictions made
in~\cite{dRHRa,dRHRb,LH15}. We thus propose a new model, whose dark
matter sector is identical to the one in the previous
model~\cite{BB14}, but whose gravitational sector is now based on
ghost-free massive bigravity theory. As bigravity theory represents
essentially a unique consistent deformation of GR, we think that the
new model will represent an important step toward a more fundamental
theory of dark matter \textit{\`a la} MOND in galactic scales. In
a separate paper~\cite{LBLH} we work out in more details the new
model and investigate whether it reproduces also the cosmological
$\Lambda$-CDM model at large scales.

\section{Dipolar Dark Matter}

A new relativistic model for dipolar dark matter was constructed
in~\cite{BB14} \textit{via} a bimetric extension of GR, which recovers
successfully the phenomenology of MOND. It relies on the existence of
two species for dark matter that couple to two different metrics and
an additional internal field in form of a vector field,
\begin{align}
\mathcal{L}&=\sqrt{-g}\biggl(\frac{M_g^2}{2}R_g-\rho_\text{b}-\rho_g\biggr)
+\sqrt{-f}\biggl(\frac{M_f^2}{2}R_f-\rho_f\biggr)
\nonumber\\ &+\sqrt{-\mathcal{G}_\text{eff}}
\biggl[M_\text{eff}^2\left(\frac{\mathcal{R}_\text{eff}}{2}
  -2\Lambda_\text{eff}\right) \nonumber\\ &\qquad\qquad\quad +
  \mathcal{A}_\mu\bigl(j_g^\mu-j_f^\mu\bigr) +
  \mathcal{W}\bigl(\mathcal{X}\bigr) \biggr]\,,\label{lagrangian}
\end{align}
where $\rho_\text{b}$, $\rho_g$, $\rho_f$ are the scalar energy
densities of pressureless ordinary matter (baryons) and the two
species of dark matter respectively, and $j_g^\mu$, $j_f^\mu$ denote
the conserved currents of the dark matter. On top of the two
Einstein-Hilbert terms for the $g$ and $f$ metrics, there is an
additional kinetic term for the effective metric
$\mathcal{G}_\text{eff}$ and a cosmological constant
$\Lambda_\text{eff}$ associated to it (here we neglect possible
cosmological constants in the $g$ and $f$ sectors). The $U(1)$ vector
field $\mathcal{A}_\mu$ is introduced to link together the two species
of dark matter particles and has a non-canonical kinetic term
$\mathcal{W}(\mathcal{X})$, with
\begin{equation}
\mathcal{X} =
\mathcal{G}_\text{eff}^{\mu\rho}\mathcal{G}_\text{eff}^{\nu\sigma}
\mathcal{F}_{\mu\nu}\mathcal{F}_{\rho\sigma}\,,
\end{equation}
and $\mathcal{F}_{\mu\nu} = \partial_\mu \mathcal{A}_\nu -
\partial_\nu \mathcal{A}_\mu$. The rich phenomenology and physical
consequences of this model were studied with great detail in
\cite{BB14}. For a particular choice of the function $\mathcal{W}$ it
recovers the desirable features of MOND and passes the constraints of
the solar system. Furthermore it agrees with the cosmological model
$\Lambda$-CDM at first order cosmological perturbation and is
  thus consistent with the fluctuations of the CMB.

The effective composite metric $\mathcal{G}_\text{eff}$ was computed
perturbatively in~\cite{BB14} and here we show the exact
non-perturbative solution for this metric. Furthermore, we investigate
the number of gravitational propagating modes and the presence of
ghost instabilities. The metric $\mathcal{G}^\text{eff}_{\mu\nu}$ was
defined in~\cite{BB14} by the implicit relations
\begin{equation}\label{relationgeff}
\mathcal{G}^\text{eff}_{\mu\nu}=\mathcal{G}_\text{eff}^{\rho\sigma}g_{\rho\mu}
f_{\nu\sigma}=\mathcal{G}_\text{eff}^{\rho\sigma}g_{\rho\nu}
f_{\mu\sigma}\,.
\end{equation}
After introducing the matrices
$G_\mu^\nu=\mathcal{G}_\text{eff}^{\nu\rho}g_{\mu\rho}$ and
$F_\mu^\nu=\mathcal{G}_\text{eff}^{\nu\rho}f_{\mu\rho}$ the above
relations simply become
\begin{equation}\label{GFinverse}
GF = FG = \Id\,,
\end{equation}
thus $G$ and $F$ are the inverse of each other. Using this fact the
form of $\mathcal{G}_\text{eff}$ can be computed. This was done
perturbatively in~\cite{BB14} with result given by~(A8)
there. Actually the solution
of~\eqref{relationgeff}--\eqref{GFinverse} can be obtained
exactly. For this, we note that the following relations are true,
\begin{equation}
g^{-1}f = \left( \mathcal{G}_\text{eff} G \right)^{-1}
\mathcal{G}_\text{eff}F = G^{-1}F = F^2\,.
\end{equation}
This means that we can identify
\begin{equation}
F = \sqrt{g^{-1}f}\,.
\end{equation}
Thus the exact solution for the effective metric fulfilling the
relations~\eqref{relationgeff} is suggestively
\begin{equation}\label{formGeff}
 \mathcal{G}^\text{eff}_{\mu\nu}=g_{\mu\rho}\Bigl(\sqrt{g^{-1}f}
 \Bigr)^\rho_\nu\,.
\end{equation}
Here we will first pay special attention to the consequences coming
from the kinetic term and a cosmological constant for this effective
metric $ \mathcal{G}_\text{eff}$ in the action~\eqref{lagrangian}.

\section{More on effective metrics}

In the different context of massive bigravity theories, interesting
proposals for an effective composite metric were made
in~\cite{dRHRa,dRHRb,LH15}. There the form of the effective metric was
determined by the question of how the coupling of the matter fields to
the two metrics of massive bigravity behave at the quantum level and
whether they alter the specific potential interactions of the allowed
potential interactions between the two metrics. One particularly
interesting effective composite metric has the following form,
\begin{equation}\label{geffdRHR}
g^\text{eff}_{\mu\nu}=\alpha^2 g_{\mu\nu} +2\alpha\beta g_{\mu\rho}
\Bigl(\sqrt{g^{-1}f}\Bigr)^{\rho}_\nu +\beta^2 f_{\mu\nu}\,,
\end{equation}
where $\alpha$ and $\beta$ are arbitrary constants. Defining the
quantities $X^{\rho}_\nu=(\sqrt{g^{-1}f})^{\rho}_\nu$ and
$Y_{\mu\nu}=g_{\mu\rho}X^{\rho}_\nu$ as was done in~\cite{dRHRa}
(where $Y_{\mu\nu}$ is shown to be symmetric), it is
straightforward to see that the determinant of this composite metric
corresponds to the allowed potential interactions in massive
bigravity,
\begin{eqnarray}
\det( g^\text{eff}_{\mu\nu})&=&\det\bigl[\left( \alpha
  g_{\mu\rho}+\beta Y_{\mu\rho}\right)g^{\rho\sigma}\left( \alpha
  g_{\nu\sigma}+\beta Y_{\nu\sigma}\right)\bigr] \nonumber\\ &=& (\det
g)^{-1}\bigl[\det(\alpha g_{\mu\nu}+\beta Y_{\mu\nu}) \bigr]^2
\nonumber \\ &=& (\det g) \bigl[\det(\alpha\Id +\beta g^{-1}Y)
  \bigr]^2\,.
\end{eqnarray}
Thus, the square root of the determinant of $g^\text{eff}_{\mu\nu}$,
say $g_\text{eff}=\det(g^\text{eff}_{\mu\nu})$, corresponds to
\begin{equation}
\sqrt{-g_\text{eff}}=\sqrt{-g} \,\det\bigl(\alpha\Id +\beta X\bigr)\,.
\end{equation}
This is the right form of the acceptable potential interactions
between the metrics $g$ and $f$. Expanding $\sqrt{-g_\text{eff}}$
around a flat background, defining
\begin{eqnarray}\label{perturbedFlat}
g_{\mu\nu} &=& (\eta_{\mu\nu}+h_{\mu\nu})^2\,, \nonumber\\ f_{\mu\nu}
&=& (\eta_{\mu\nu}+\ell_{\mu\nu})^2\,,
\end{eqnarray}
they correspond to the specific interactions of the form
\begin{equation}\label{detgeff}
\sqrt{-g_\text{eff}} = \sum_{n=0}^4 (\alpha+\beta)^{4-n} e_n(k)\,,
\end{equation}
where $k_{\mu\nu}=\alpha h_{\mu\nu} + \beta \ell_{\mu\nu}$, and the
symmetric polynomials are defined by (with $[\cdots]$ denoting the
trace as usual)
\begin{eqnarray}\label{polynomials}
e_0(k)&=&1\,, \nonumber\\ e_1(k)&=& \bigl[k\bigr]\,,
\nonumber\\ e_2(k)&=&
\frac{1}{2}\bigl(\bigl[k\bigr]^2-\bigl[k^2\bigr]\bigr)\,,
\nonumber\\ e_3(k)&=&
\frac{1}{6}\bigl(\bigl[k\bigr]^3-3\bigl[k\bigr]\bigl[k^2\bigr]
+2\bigl[k^3\bigr]\bigr)\,, \nonumber\\ e_4(k)&=&
\frac{1}{24}\bigl(\bigl[k\bigr]^4-6\bigl[k\bigr]^2\bigl[k^2\bigr]
+3\bigl[k^2\bigr]^2 \nonumber\\ &&\quad +
8\bigl[k\bigr]\bigl[k^3\bigr]-6\bigl[k^4\bigr]\bigr)\,.
\end{eqnarray}
Thus, $\sqrt{-g_\text{eff}}$ has exactly the nice structure of the
potential with the special tuning in order to remove the BD ghost at
any order~\cite{deRham10, dRGT10}.

Finally, the relation between the effective composite metric
$\mathcal{G}_\text{eff}$ proposed in~\cite{BB14} and the alternative
effective metric $g_\text{eff}$ proposed in~\cite{dRHRa} is
given by [since $\mathcal{G}^\text{eff}_{\mu\nu}=Y_{\mu\nu}$
  from~\eqref{formGeff}]
\begin{equation}
g^\text{eff}_{\mu\nu}=\alpha^2 g_{\mu\nu}+2\alpha\beta
\,\mathcal{G}^\text{eff}_{\mu\nu}+\beta^2 f_{\mu\nu}\,.
\end{equation}
In other words, $\mathcal{G}_\text{eff}=g\sqrt{g^{-1}f}$ does not
contain the linear parts proportional to $g$ and $f$ in
$g_\text{eff}$. Unfortunately, this will have important consequences
as we will see in the following section.

\section{Cosmological constant for the effective metric} 

We will first study the consequences of having in the
model~\eqref{lagrangian} the square root of the determinant
$\mathcal{G}_\text{eff}=\det(\mathcal{G}^\text{eff}_{\mu\nu})$. Since
our non-perturbative solution is
$\mathcal{G}^\text{eff}_{\mu\nu}=g_{\mu\rho}(\sqrt{g^{-1}f})^\rho_\nu$,
the square root of the determinant reads
\begin{equation}
\sqrt{-\mathcal{G}_\text{eff}} = \sqrt{\sqrt{-g}\sqrt{-f}}\,.
\end{equation}
Perturbed around a flat background, it corresponds in the
notation~\eqref{perturbedFlat} to
\begin{align}
\sqrt{-\mathcal{G}_\text{eff}} &= 1+\frac{1}{2}\bigl[h + \ell\bigr] +
\frac{1}{8}\Bigl(\bigl[h+\ell\bigr]^2
-2\bigl[h^2+\ell^2\bigr]\Bigr)\nonumber\\ &+
\!\!\frac{1}{48}\Bigl(\bigl[h+\ell\bigr]^3
-6\bigl[h+\ell\bigr]\bigl[h^2+\ell^2]+8\bigl[h^3+\ell^3\bigr]\Bigr)
\nonumber\\ &+ \cdots \,.
\end{align}
As is immediately seen, $\sqrt{-\mathcal{G}_\text{eff}}$ does not have
the right potential structure, in fact it does not even contain the
right structure for the linear Fierz-Pauli mass term. Any Lagrangian
that contains this term as a possible potential interaction between
the two metrics has immediately the BD ghost at the linear order. Thus
the cosmological constant for this effective metric or any minimal
coupling to matter fields \textit{via}
$\mathcal{G}^\text{eff}_{\mu\nu}$ will reintroduce the dangerous
ghostly mode. The ghost would come already at a scale
\begin{equation}
  m^2\mpl^2\sqrt{-\mathcal{G}_\text{eff}}\sim \frac{m^2 \mpl^2(\Box\pi)^2}{\Lambda_3^6}= \frac{(\Box\pi)^2}{m^2}\,,
\end{equation}
where $\Lambda_3^3=\mpl m^2$ and $\pi$ denotes the 0-helicity
mode. This means that the ghost is a very light degree of
freedom. This immediately kills the possibility of considering any
Lagrangian (independently of all the additional terms present in it) that
contains $\sqrt{-\mathcal{G}_\text{eff}}$.

\section{Mini-superspace of the new kinetic term}

In the previous section we studied the implications of having the
cosmological constant for $\mathcal{G}^\text{eff}_{\mu\nu}$ and saw
that it introduces ghostly interactions between the two metrics. In
this section we will pay attention to the kinetic term
$\sqrt{-\mathcal{G}_\text{eff}} \mathcal{R}_\text{eff}$, where
$\mathcal{R}_\text{eff}$ is the Ricci scalar built from
$\mathcal{G}^\text{eff}_{\mu\nu}$. Moreover, we will investigate the
allowed number of kinetic terms. The first test that such term has to
pass is the special case of the mini-superspace. The respective
metrics in the mini-superspace are given by
\begin{eqnarray}
\ud s_g^2&=&g_{\mu\nu} \ud x^\mu \ud x^\nu = -n_g^2 \ud t^2 + a_g^2
\ud x^2\,, \nonumber\\ \ud s_f^2&=&f_{\mu\nu} \ud x^\mu \ud x^\nu =
-n_f^2 \ud t^2 + a_f^2 \ud x^2\,,
\end{eqnarray}
where $n_g$, $n_f$ and $a_g$, $a_f$ are functions of the cosmic
time $t$ only. Consider the following Lagrangian with the three
kinetic terms
\begin{equation}
\mathcal{L}^\text{eff}_\text{kin} = \frac{M_g^2}{2} \sqrt{-g} R_g +
\frac{M_f^2}{2} \sqrt{-f} R_f +
\frac{M_\text{eff}^2}{2}\sqrt{-\mathcal{G}^\text{eff}}
\mathcal{R}_\text{eff}\,,
\end{equation}
that in the mini-superspace simply becomes
\begin{equation}
\mathcal{L}^\text{eff}_\text{kin} = -\frac{3M_g^2 a_g \dot{a}_g^2}{n_g} -
\frac{3 M_f^2 a_f \dot{a}_f^2}{n_f} - \frac{3M_\text{eff}^2
  a_\text{eff} \dot{a}_\text{eff}^2}{n_\text{eff}}\,.
\end{equation}
Following the prescription~\eqref{relationgeff} we obtain
$n_\text{eff}=\sqrt{n_gn_f}$ and $a_\text{eff}=\sqrt{a_ga_f}$. We
compute the conjugate momenta for the scale factors and get
\begin{align}
p_g &=-6M_g^2a_g^2H_g-\frac{3}{2}M_\text{eff}^2a_f
\sqrt{\frac{a_ga_f}{n_gn_f}}\bigl(H_gn_g+H_fn_f\bigr)\,,
\nonumber\\ p_f &=-6M_f^2a_f^2H_f-\frac{3}{2}M_\text{eff}^2a_g
\sqrt{\frac{a_ga_f}{n_gn_f}}\bigl(H_gn_g+H_fn_f\bigr)\,, 
\end{align}
where $H_g=\frac{\dot{a}_g}{a_gn_g}$ and similarly $H_f$ are the
conformal Hubble factors. Now we can perform the Legendre
transformation to obtain the following Hamiltonian:
\begin{align}\label{hamiltonian}
& \mathcal{H}^\text{eff}_\text{kin} = \frac{1}{\mathcal{Q}}\Bigl\{
  a_g^2n_f\Bigl(M_\text{eff}^2p_{g}^2\sqrt{a_g a_f}n_g+4M_g^2
  p_{f}^2a_f\sqrt{n_g n_f}\Bigr) \nonumber\\& \quad + 2 p_{g} a_g a_f
  n_g \Bigl(-M_\text{eff}^2
  p_{f}\sqrt{a_ga_f}n_f+2M_f^2p_{g}a_f\sqrt{n_gn_f}\Bigr) \nonumber
  \\ & \quad + M_\text{eff}^2p_{f}^2a_f^2\sqrt{a_ga_f}n_gn_f
  \Bigr\}\,,
\end{align}
where we defined the shortcut notation for convenience 
\begin{eqnarray}
\mathcal{Q}&=&-12\Bigl(M_\text{eff}^2M_f^2a_f^3\sqrt{a_ga_f}n_g+M_\text{eff}^2
M_g^2a_g^3\sqrt{a_ga_f}n_f \nonumber\\&&\qquad
+4M_f^2M_g^2a_g^2a_f^2\sqrt{n_g n_f}\Bigr)\,.
\end{eqnarray}
The Hamiltonian is highly non-linear in the lapses $n_g$ and
$n_f$. Since there is no shift over which we have to integrate, this
is an immediate sign that these three kinetic terms have the BD ghost
degree of freedom already in the mini-superspace (see
\cite{GaugeFields} for an introduction to constrained hamiltonian
systems). Thus, one has to avoid the two very bad contributions in
form of (i) the cosmological constant term for
$\mathcal{G}_\text{eff}$, and (ii) the kinetic term
$\sqrt{-\mathcal{G}_\text{eff}} \mathcal{R}_\text{eff}$ --- both these
terms correspond to ghostly interactions. Because of their very
different structures there is no hope for cancellations between these
terms.

Taking the limit when $M_f \to 0$ of the Hamiltonian
\eqref{hamiltonian} results in
\begin{align}
\mathcal{H}^\text{eff}_\text{kin}\big|_{M_f\to0} =&
-\frac{1}{12a_g^3}\biggl( \frac{(p_{g}a_g-p_{f}a_f)^2n_g}{M_g^2}
\nonumber\\&\qquad\quad\quad +\frac{4p_{f}^2a_g\sqrt{a_g
    a_f}\sqrt{n_gn_f}}{M_\text{eff}^2}\biggr)\,.
\end{align}
As one can see, even in this limit the Hamiltonian is not linear
  in the lapses, so that the variation of the Hamiltonian with
respect to the lapses gives rise to equations of motion that depend on
the lapses and hence the constraint equation is lost. Therefore, the
kinetic term $\sqrt{-\mathcal{G}_\text{eff}} \mathcal{R}_\text{eff}$
introduces ghostly interactions already in the mini-superspace
independently of the number of present kinetic terms.

An interesting question to address at this stage is whether or not the
mini-superspace can be made ghost-free by considering the kinetic term
for $g_\text{eff}$ that was proposed in~\cite{dRHRa}. Since
the determinant of $g_\text{eff}$ corresponds to the right ghost-free
potential interactions between two metrics, the kinetic term for
$g_\text{eff}$ might behave better than that for
$\mathcal{G}_\text{eff}$. Thus, consider as next the Lagrangian with
the alternative three kinetic terms
\begin{equation}
\tilde{\mathcal{L}}^\text{eff}_\text{kin} = \frac{M_g^2}{2} \sqrt{-g}
R_g + \frac{M_f^2}{2} \sqrt{-f} R_f +
\frac{M_\text{eff}^2}{2}\sqrt{-g_\text{eff}} R_\text{eff}\,,
\end{equation}
where $R_\text{eff}$ is now the Ricci scalar of the metric
$g^\text{eff}_{\mu\nu}$. In the mini-superspace this becomes
\begin{equation}
\tilde{\mathcal{L}}^\text{eff}_\text{kin} = -\frac{3M_g^2 a_g
  \dot{a}_g^2}{n_g} - \frac{3 M_f^2 a_f \dot{a}_f^2}{n_f} -
\frac{3M_\text{eff}^2 \tilde{a}_\text{eff}
  \dot{\tilde{a}}_\text{eff}^2}{\tilde{n}_\text{eff}}\,,
\end{equation}
with this time $\tilde{n}_\text{eff}=\alpha n_g +\beta n_f$ and
$\tilde{a}_\text{eff}=\alpha a_g + \beta a_f$. The conjugate momenta for
the scale factors are now
\begin{align}
\tilde{p}_{g} &= -6\Bigl(M_g^2a_g^2H_g+\frac{\alpha M_\text{eff}^2
  \tilde{a}_\text{eff}}{\tilde{n}_\text{eff}}\bigl(\alpha
a_gH_gn_g+\beta a_fH_fn_f\bigr)\Bigr)\,,\nonumber\\ \tilde{p}_{f} &=
-6\Bigl(M_f^2a_f^2H_f+\frac{\beta M_\text{eff}^2
  \tilde{a}_\text{eff}}{\tilde{n}_\text{eff}}\bigl(\alpha
a_gH_gn_g+\beta a_fH_fn_f\bigr)\Bigr)\,.
\end{align}
Thus, the Hamiltonian is given by
\begin{align}\label{hamilton_geff}
& \tilde{\mathcal{H}}^\text{eff}_\text{kin} =
  \frac{1}{\tilde{\mathcal{Q}}}\Bigl\{
  a_gn_f\bigl[-\alpha(M_g^2\tilde{p}_{f}^2+M_\text{eff}^2(\alpha\tilde{p}_{f}
    - \beta\tilde{p}_{g})^2)n_g \nonumber\\& \quad
    -M_g^2\tilde{p}_{f}^2\beta n_f\bigr] +a_f
  n_g\bigl[-M_f^2\tilde{p}_{g}^2\alpha n_g-\beta(M_f^2\tilde{p}_{g}^2
    \nonumber\\& \quad +M_\text{eff}^2(\alpha\tilde{p}_{f} -
    \beta\tilde{p}_{g})^2)n_f\bigr] \Bigr\}\,,
\end{align}
where $\tilde{\mathcal{Q}}$ stands for
\begin{eqnarray}
\tilde{\mathcal{Q}}&=&12\Bigl(M_f^2\alpha
a_f\bigl[(M_g^2+M_\text{eff}^2\alpha^2)a_g+M_\text{eff}^2\alpha\beta
  a_f\bigr]n_g \nonumber\\&&\quad +M_g^2\beta a_g
\bigl[M_f^2a_f+M_\text{eff}^2\beta
  \tilde{a}_\text{eff}\bigr]n_f\Bigr)\,.
\end{eqnarray}
Again, the Hamiltonian is highly non-linear in the lapses.
The problem comes from the fact that we have too many kinetic
terms. Indeed we see immediately that the only way of having linear
dependence in the lapses in the mini-superspace (and hence getting rid
of the BD ghost) is if we take either the $\beta\to 0$ limit --- this
would simply correspond to having only the standard kinetic terms for
the $g$ and $f$ metrics --- or the $M_f\to 0$ limit,
\begin{align}
&\tilde{\mathcal{H}}^\text{eff}_\text{kin}\big|_{M_f \to 0}= -
  \frac{M_\text{eff}^2\beta(\alpha \tilde{p}_{f} - \beta
    \tilde{p}_{g})^2a_f
    n_g}{12M_\text{eff}^2M_g^2\beta^2a_g\tilde{a}_\text{eff}}
  \nonumber\\& - \frac{\alpha\bigl(M_g^2
    \tilde{p}_{f}^2+M_\text{eff}^2(\alpha \tilde{p}_{f} -\beta
    \tilde{p}_{g})^2\bigr)n_g+M_g^2\tilde{p}_{f}^2\beta
    n_f}{12M_\text{eff}^2M_g^2\beta^2\tilde{a}_\text{eff}}\,.\label{limitMf0}
\end{align}
As we see, the Hamiltonian becomes linear in the lapses when we remove
for instance the kinetic term for the $f$ metric. Thus, the only way
of having a healthy mini-superspace is if we restrict the kinetic
Lagrangian to be either $M_g^2 \sqrt{-g} R_g + M_f^2\sqrt{-f} R_{f}$
which are the standard ghost-free kinetic terms, or $M_g^2
\sqrt{-g} R_g + M_\text{eff}^2\sqrt{-g_\text{eff}} R_\text{eff}$. In a
symmetric manner we could also remove the kinetic term for $g$ and
hence $M_f^2 \sqrt{-f} R_f + M_\text{eff}^2\sqrt{-g_\text{eff}}
R_\text{eff}$ would be also perfectly valid. In summary, one should
restrict the theory to have not more than two kinetic terms in order
not to reintroduce the BD ghost.

\section{Dipolar Dark Matter in ghost-free bimetric theory}

The dark matter model proposed in~\cite{BB14} is therefore
non-viable, but nevertheless points toward an interesting connection
between dark matter at small galactic scales (interpreted as
DDM) and bimetric gravity. Based on our previous analysis, we
would like now to propose the following new model for dipolar dark
matter based on ghost-free bimetric theory,
\begin{align}
&\!\!\!\mathcal{L}_\text{new} =
  \sqrt{-g}\biggl(\frac{M_g^2}{2}R_g -\rho_\text{b}-\rho_g\biggr)
  +\sqrt{-f}\biggl(\frac{M_f^2}{2}R_f-\rho_f\biggr)
  \nonumber\\ &\qquad +\sqrt{-g_\text{eff}} \biggl[m^2 M_\text{eff}^2
    + \mathcal{A}_\mu\bigl(j_g^\mu-j_f^\mu\bigr) +
    \mathcal{W}\bigl(X\bigr) \biggr]\,,\label{newlagrangian}
\end{align}
where the ghost-free potential interactions are defined by the metric
\eqref{geffdRHR} [they take the form
\eqref{detgeff}--\eqref{polynomials} when expanded around a flat
background], and where the kinetic term of the vector field is
now constructed with the metric $g^\text{eff}_{\mu\nu}$,
\begin{equation}
X = g_\text{eff}^{\mu\rho}
g_\text{eff}^{\nu\sigma}\mathcal{F}_{\mu\nu}\mathcal{F}_{\rho\sigma}\,.
\end{equation}
As was shown in~\cite{dRHRa,dRHRb,LH15}, the matter fields can
separately couple to either the $g$ metric or $f$ metric without
invoking the BD ghost. Additionally the matter fields can couple to
the effective composite metric $g_\text{eff}$ which is ghost-free in
the mini-superspace and in the decoupling limit. 

Here we propose to couple the ordinary baryonic fields with mass
density $\rho_\text{b}$ to the standard $g$ metric while coupling the
two species of dark matter with densities $\rho_g$ and $\rho_f$
separately to the $g$ and $f$ metrics respectively. Furthermore, in
order to link together the two species of dark matter particles, we
consider a vector field $\mathcal{A}_\mu$ that minimally couples to
the effective metric $g_\text{eff}$. The vector field plays the role
of a ``graviphoton'' since it is generated by the mass currents
$j_g^\mu$ and $j_f^\mu$ of the particles. The presence of this
internal field is necessary to stabilize the dipolar medium and is
expected to yield the wanted mechanism of polarization. 

The model~\eqref{newlagrangian} fulfils the restrictions coming from
our previous analysis, as it contains no more than two kinetic terms
(in particular the problematic kinetic term for
$\mathcal{G}_\text{eff}$ is absent), and the potential interactions
between the two metrics coming from the square root of the determinant
of $g_\text{eff}$ correspond to the ghost-free prescriptions.

However, one needs also to be careful with the assumptions on the
  dark matter fields and their currents. Indeed the ghost could still
  be present in the matter sector.\footnote{We are grateful to
      Claudia de Rham for discussions on this.} Since the dark matter
  fluids that live on the $g$ and $f$ metrics directly couple to
  $A_\mu$ that lives on the $g_\text{eff}$ metric, there is \textit{a
    priori} the danger of having the ghost present due to the
  interaction term in the matter sector. We will investigate this
  question with great detail in \cite{LBLH} and see if for a specific choice
  of the dark matter fields the ghost can be maintained absent. For this
  we shall perform the decoupling limit analysis of our new model \eqref{newlagrangian}
  including the matter sector and study the required amount of initial 
  conditions.

Finally, the model \eqref{newlagrangian} should share the nice
properties and phenomenology of the model proposed in~\cite{BB14} and
therefore provides a promising road for a relativistic dipolar dark
matter model to be investigated. The MOND phenomenology, the PPN
parameters and the cosmology of this model will be studied in a
separate paper~\cite{LBLH}.

\section{Conclusions}

We explored the possible candidates for relativistic dark matter
models in bimetric extensions of General Relativity, that hopefully
will provide modified Newtonian dynamics (MOND) at galactic scales
while giving rise to an expansion at cosmological scales. A promising
road comes from the ghost-free constructions of dRGT massive
gravity~\cite{deRham10, dRGT10} where the interactions between two
metrics are tuned in a way that the Boulware-Deser ghost remains
absent. Furthermore, the important studies of possible consistent
couplings to matter fields~\cite{dRHRa,dRHRb,LH15} are beneficial to
us, since for the model to work, we have to consider two different
species of dark matter particles that couple separately to the two
metrics while an additional internal vector field couples minimally to
an effective metric built out of the two. The vector field links
together the two sectors of the dark matter particles and plays a
crucial role for gravitational polarization and MOND~\cite{BB14,
  BBwag}.

For the ghost absence the question of allowed kinetic interactions is
mandatory. We showed that the kinetic Lagrangian containing three
kinetic terms immediately gives rise to the introduction of the ghost
and we therefore concluded that only two kinetic terms are allowed. 

In a future work~\cite{LBLH}, we will study in detail the covariant
equations of motion of the new model, derive the non-relativistic
limit and see if the polarization mechanism for dark matter works in
the same way as in the originally proposed model. We will investigate
in detail the possible danger of ghostly interactions in the matter sector
and constrain further the model. We intend also to
check if the parametrized post-Newtonian parameters are close to the
ones of GR in the solar system, and to investigate the cosmological
solutions in first order perturbations.

\acknowledgments We would like to thank C\'edric Deffayet, Claudia de
Rham, Gilles Esposito-Far\`ese, Nima Khosravi and Shinji
Mukohyama for very useful and enlightening discussions. L.H. wishes to
acknowledge the Institut d'Astrophysique de Paris for hospitality and
support at the initial stage of this work.

\bibliography{ListeRefLH.bib}

\end{document}